\documentclass[11pt]{article}
\usepackage[top=1in, bottom=1in, left=1in, right=1in]{geometry}

\usepackage{framed}
\usepackage{etoolbox}
\usepackage{subcaption}
\usepackage{amsmath}
\usepackage{subcaption}
\usepackage{mathtools}  
\usepackage{enumitem}
\usepackage[colorlinks=true, citecolor=brown]{hyperref}
\usepackage[numbers]{natbib}
\usepackage{url}
\usepackage{booktabs}

\usepackage{parskip}
\usepackage{mathpazo}

\usepackage{bibentry}

\usepackage{amssymb}
\usepackage{amsmath}

\usepackage{xcolor}
\usepackage{booktabs}

\usepackage{nicefrac}
\usepackage{subcaption}
\usepackage{authblk}

\newcommand{\E}{\mathbb{E}}

\newcommand{\X}{\mathcal{X}}

\newcommand{\set}[1]{\{#1\}}

\newcommand{\norm}[1]{\left|\left|#1\right|\right|}
\newcommand{\ccolor}[1]{{\leavevmode\color{black}#1}}

\newcommand{\abs}[1]{\left|#1\right|}

\title{Moral Change or Noise? On Problems of Aligning AI With Temporally Unstable Human Feedback}

\author[1]{Vijay Keswani}
\author[1]{Cyrus Cousins}
\author[1]{Breanna Nguyen}
\author[2]{Vincent Conitzer$^*$}
\author[2]{Hoda Heidari$^*$}
\author[1]{Jana Schaich Borg$^*$}
\author[1]{Walter Sinnott-Armstrong$^*$}

\affil[1]{Duke University}
\affil[2]{Carnegie Mellon University}

\date{}

\begin{document}

\maketitle
\def\thefootnote{*}\footnotetext{These authors contributed equally and are listed alphabetically.}\def\thefootnote{\arabic{footnote}}

\begin{abstract}
Alignment methods in moral domains seek to elicit moral preferences of human stakeholders and incorporate them into AI. This presupposes moral preferences as static targets, but such preferences often evolve over time. Proper alignment of AI to dynamic human preferences should ideally account for ``legitimate'' changes to moral reasoning, while ignoring changes related to attention deficits, cognitive biases, or other arbitrary factors. However, common AI alignment approaches largely neglect temporal changes in preferences, posing serious challenges to proper alignment, especially in high-stakes applications of AI, e.g., in healthcare domains, where misalignment can jeopardize the trustworthiness of the system and yield serious individual and societal harms.  This work investigates the extent to which people’s moral preferences change over time, and the impact of such changes on AI alignment. Our study is grounded in the \emph{kidney allocation} domain, where we elicit responses to pairwise comparisons of hypothetical kidney transplant patients from over 400 participants across $3$--$5$ \ccolor{sessions}. We find that, on average, participants change their response to the same scenario presented at different times around 6--20\% of the time (exhibiting ``response instability''). Additionally, we observe significant shifts in several participants’ retrofitted decision-making models over time (capturing ``model instability''). The predictive performance of simple AI models decreases as a function of both response and model instability. Moreover, predictive performance diminishes over time, highlighting the importance of accounting for temporal changes in preferences during training. These findings raise fundamental normative and technical challenges relevant to AI alignment, highlighting the need to better understand the object of alignment (what to align to) when user preferences change significantly over time, including the mechanisms underlying these changes.

\end{abstract}

\section{Introduction}

Alignment of modern AI systems to human preferences has emerged as a cornerstone of the current pursuit of responsible, ethical, and safe AI. 
Various failure modes of AI --- such as biased behavior \cite{santurkar2023whose}, reward hacking \cite{pan2022effects}, and even existential threats \cite{shevlane2023model} --- have been framed as stemming from misalignment between AI's goals and the values and preferences of its stakeholders.
As a result, the field of AI alignment aspires to address these issues by ensuring that AI outputs are guided (and often constrained) by user and societal preferences
\cite{ji2023ai}.  
The advertised applicability of this approach in encoding \textit{subjective} preferences 
in AI systems has also led to its proposed usage in \textit{moral domains}
\cite{kim2018computational, noothigattu_voting-based_2018, johnston2023deploying,tennant2024moral}.
Since moral dilemmas are bound to arise in several consequential applications of AI --- e.g., designing equitable healthcare resource allocation policies \cite{sinnott2021ai} or characterizing autonomous vehicle behavior \cite{awad2018moral} --- aligning the AI to the moral preferences of the user or the aggregated preferences of a community provides a pathway towards encoding 
ethical values in AI's behavior.

The standard approach to building morally-aligned AI involves a one-time elicitation of moral preferences from users \cite{awad2018moral,freedman2020adapting,johnston2023deploying}.
Yet, human preferences are dynamic, evolving over time and across contexts \cite{tversky1981framing}.
Cognitive processes underlying moral judgments also fluctuate, reflecting changes in values informing moral decisions, updated moral reasoning in light of new information, or even alterations in time-on-task and cognitive capacity
\cite{amir2008choice, warren2011values, jia2022effects}.
However, much of the existing work in AI alignment for moral domains neglects temporal changes to the underlying moral preferences of stakeholders, the consequences of which can be severe in real-world applications \cite{dung2023current}.
The possibility of negative impacts of temporal misalignment and the increased use of AI for decision-making have indeed led to several recent calls for real-world data collection to assess the impact of preference changes on AI alignment \cite{yehposition, boerstler2024stability, carroll2024ai}.
Our work answers this call with a data-driven investigation into dynamic moral preferences.

\paragraph{Our Contributions.}
We investigate the AI alignment challenges associated with temporal changes in moral preferences. 
Our work studies moral preferences related to kidney transplant allocation, where advocates have noted the ability of AI to improve transplant efficiency \cite{schwantes2021technology} and align transplant decisions with stakeholder preferences \cite{freedman2020adapting}.

Simulating a setting where kidney patients outnumber available kidneys, each participant in our study is presented with several pairwise comparisons of hypothetical kidney patients and asked to choose which patient should receive the kidney if only one is available.
In this morally high-stakes domain, we elicit preferences of more than 400 survey participants over three to five sessions, spanning up to two weeks.
In each session, participants are presented with 60 pairwise comparisons, with six comparisons repeated across all sessions and twice per session (Section~\ref{sec:study_methodology}).

Participants' responses to the repeated scenario allow us to measure their ``response instability,'' i.e., how often they change their response to the same scenario repeated at different times/during different sessions.
We observed that response instability, on average, varies from 6--20\%, with significantly higher instability for scenarios that were relatively more challenging (i.e., containing \textit{tradeoffs} across several morally-relevant factors; Section~\ref{sec:observations_response_instability}).
We also investigate the extent to which participants seem to change their decision-making processes over time.
Evaluation of {agreement} between models learned from participants' session-wise responses shows that the overlap between predictions from session-wise models decreases with time;
we call this phenomenon ``model instability'' (Section~\ref{sec:observations_model_instability}).
To understand the reasons for changing preferences, we categorize participants by their response stability and model stability levels, and study variations in decision-making properties across participant categories.
We provide evidence that different participants exhibit different mechanisms of preference change (Section~\ref{sec:observations_alignment}).
For some, preference change arises from a reduction in model complexity (e.g., reducing the number of factors they account for in making their decisions) over time.
For others, instability reflects stochastic changes in their decision processes (which may be due to morally arbitrary factors, such as attention, time of day, and beyond).
Crucially, different mechanisms underlying the change in preferences prescribe different approaches to AI alignment. While it is essential for alignment methods to identify and capture legitimate changes to moral preferences of stakeholders (e.g., those resulting from participants' judgment evolving with more exposure and reflection), morally arbitrary changes, such as noisy responses due to fatigue and attention fluctuations, should not impact alignment.  

To assess the impact of preference change on AI alignment, we train preference optimization models on participant- and population-level data.
We observe a significant association between temporal instability and the trained model's error rate in predicting participant responses
(Section~\ref{sec:observations_alignment}).
In particular, for participants who exhibited high levels of response and model instability, predictive error rate was higher by 5-16\%, in comparison to those who were response and model stable.
Additionally, a temporal analysis shows that error rates of trained models increase over time for participants who were response and model unstable.

Overall, our study highlights several challenges and corresponding opportunities for AI alignment.
Our findings emphasize a prominent normative challenge that needs to be tackled by AI alignment methods: When human moral preferences change over time, with which should AI align? 
The earlier preference, or the later one? \emph{Neither?} Perhaps both, their average, \emph{or something else entirely?} 
The answer is nuanced, depending on whether the change is arbitrary or involves adopting new values.
\ccolor{Additionally, our work highlights fundamental deficiencies of choice-based preference alignment methods in deciphering how and why people change their moral preferences, and presents opportunities to build better methods that are cognizant of the mechanisms of preference change (Section~\ref{sec:discussion}).}

\section{Related Work}

Recent surges in the use of AI, and the associated risks, have led to increased research on AI alignment.
We refer the reader to \citet{ji2023ai} and \citet{shen2024towards} for an overview of AI alignment methods.
While these methods generally rely on one-shot/episode collection of human feedback, human preferences are dynamic and context-sensitive \citep{slovic1995construction,zhi-xuan_beyond_2024}. 
Recent works have noted many issues with assuming static preferences, such as conflicts between learned and desired preferences \citep{carroll2024ai, curmei2022towards, kleinberg2024inversion} and flattening of preference heterogeneity \citep{buyl2025ai}.
These theoretical challenges motivate our 
empirical study of human preferences over time.
Temporal misalignment is
especially pressing for preference optimization methods that train models on fixed datasets of human choices \citep{liu_survey_2025, boerstler2024stability}. Recent studies highlight that models trained on such data generalize poorly under distributional shift \citep{lin_limited_2024}, and that aggregating heterogeneous preferences can lead to inconsistent performance \citep{shirali_direct_2025}.
Such findings cast doubt on the reliability of one-shot preference elicitation.

A wide range of alignment research uses moral preference datasets
as foundational benchmarks. The Moral Machine and ETHICS datasets, for example, have informed the development and evaluation of models for moral reasoning across various domains \citep{noothigattu_voting-based_2018, wiedeman_modeling_2020, rodionov_evaluation_2023, hendrycks_aligning_2023, zaim_bin_ahmad_large-scale_2025}. Similar strategies are applied in medical ethics and resource triage settings \citep{dickerson2025gets, mohsin_learning_2025}. 
Yet, AI moral alignment methods often treat human preferences as static, despite evidence from moral psychology showing otherwise \cite{rehren2023stable}. 
\citet{boerstler2024stability} investigated the scale of moral judgment instability in the kidney allocation domain; however, their studies had relatively small participant pools (fewer than 50) and limited analysis of impact of instability on AI alignment.
Our study follows a similar design, but with significantly more participants to better assess the impact of temporal instabilities on alignment. 
The consequences of misalignment 
can be much more severe in moral AI domains.
If people change their movie/music preferences over time, a misaligned recommender AI would 
result in unmet entertainment needs.
In contrast, when someone has one medical resource allocation policy one day but changes it to create a more equitable policy on a later day, 
then misalignment in this moral domain can be seen to be much more consequential. 
Our work is motivated by the need to tackle these potential consequences of temporal misalignment.

\section{Study Methodology} \label{sec:study_methodology}

\subsection{Data Collection}
Participants were recruited using Qualtrics
and asked to take part in {five sessions}, with up to three days between sessions.
Each session sought their judgments for several moral scenarios, consisting of pairwise comparisons between two kidney patients.
Participants were asked to ``{choose which of two patients--Patient A or Patient B--should receive priority for a kidney when only one is available}.''

\paragraph{Scenario Design.}
In each kidney allocation scenario, hypothetical patients A and B are described using the following $d{=}8$ features: (a) \textit{number of child dependents}, (b) \textit{life years to be gained due to a successful kidney transplant}, (c) \textit{number of alcoholic drinks per day before diagnosis}, (d) \textit{number of past violent crimes}, (e) \textit{obesity level}, (f) \textit{hours per week patient is expected to be able to work after kidney transplant}, (g) \textit{years on the transplant waiting list}, and (h) \textit{chance of patient's immune system rejecting transplanted kidney}.
All features were selected based on a review of transplantation practices and prior studies \cite{transplantliving,keswani2025can}, and
included both medical and non-medical attributes that people considered morally relevant. 
Appendix~\ref{sec:methodology_appendix} presents an example scenario and additional feature descriptions.

\noindent
\textbf{Repeated Scenarios.}
To assess whether participants provided the same response for a scenario presented at different times, six scenarios were repeated, both across all sessions, and twice within each session.
These repeated scenarios were designed to vary in \textit{expected difficulty} by varying the number of tradeoffs between Patients A and B.
Scenarios $U_1$ and $U_2$ have the fewest tradeoffs, with one feature favoring one patient (A or B) and the remaining seven features favoring the other.
Scenarios $V_1$ and $V_2$ add further tradeoffs, with two features favoring A, two favoring B, and the remaining four features 
equal for both.
Scenarios $W_1$ and $W_2$ then maximize tradeoffs, differing across all features, with four features favoring A and the remaining four favoring B.
\ccolor{Presentation order for all scenarios was randomized each time, both in terms of feature order and which patient appeared left vs.\ right, to prevent participants from remembering them.}

These repeated scenarios were selected from a larger pool that was tested in a pilot study, where the 
chosen scenarios displayed a wide range of across-participant disagreement.
We also included two attention-check scenarios each session.
Complete descriptions of repeated scenarios are presented in Appendix~\ref{sec:methodology_appendix}, along with pilot study details.
For all other scenarios, the features for both patients were \emph{independently} and \emph{uniformly randomly} sampled from their domains.

\paragraph{Participants.}
Overall, 1410 participants took part in this study. However, not all completed all five sessions, and several failed attention check questions.
Excluding those who did not pass all attention checks left 1227 participants.
Among these, 132 completed five sessions, 318 
completed at least four sessions, and 404 
completed at least three sessions. 
We present only results on 
this cohort 
who provided valid responses across at least three sessions. 
\ccolor{All participants were compensated at the rate of \$12/hr. 
The survey methodology was approved by the Institutional Review Board (IRB) of the first author's institution. Additional preprocessing steps and aggregate demographics are
noted in Appendix~\ref{sec:methodology_appendix}.
}

\paragraph{Qualitative Responses.}
We also asked participants who completed all five sessions to self-report their decision process at the end of the last session, by scoring and ranking the importance of all features,
and by textually describing their decision strategy.
Deviations of \textit{learned} vs.\ \textit{self-reported} preferences are documented in Appendix~\ref{sec:other_instability_results}.

\subsection{Metrics}

Our study aims to assess temporal shifts in moral preferences.
Such shifts can manifest themselves in the form of \textit{unsteady} judgments 
and/or differences in models learned using participants' judgments across sessions.
To quantify these changes, we therefore measure both \textit{response stability} and \textit{model stability}.
Additionally, we define decision-making properties that might explain temporal instabilities, e.g., \textit{scenario difficulty}, \textit{model shift}, and \textit{model complexity}.

\begin{figure}
    \centering
    \includegraphics[width=0.7\linewidth]{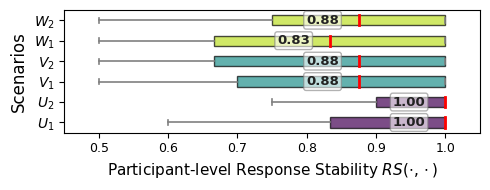}
    \caption{\ccolor{Response stability distribution (median annotated) for all repeated scenarios. Participants were relatively more stable for $U_1, U_2$ compared to other scenarios.}}
    \label{fig:agreement_and_stability}
\end{figure}

\paragraph{Response Stability.}
We first measure and report how \textit{stable} participants were for the repeated scenarios.
For any repeated scenario $S \in \set{U_1, U_2, V_1, V_2, W_1, W_2}$,
the \textit{dominant response} is the patient chosen by participant $i$
50\% or more of the time for $S$, and the 
\emph{response stability} is the fraction of responses that deviate from the dominant response, and thus falls in $[0.5, 1]$.
Formally, we define the response stability of participant $i$ for scenario $S$ as
\[ \textrm{RS}(S; i) := \frac{\#\text{times $i$ chose \textit{dominant response} for } S}{\text{total \#times $S$ was presented to participant } i}.\]
We also define \textit{average response stability} 
over repeated scenarios as $\text{avgRS}(i)=$  ${\frac{1}{6}} \sum_{S \in \set{U_1, U_2, \ldots}} \text{RS}(S; i)$.

\paragraph{Model Stability.}
To understand if and how participants change their decision process over time, we quantify the extent of \textit{agreement} between predictions made by models learned from participant responses in each session.
For any participant $i$ and session $j$, we use the participant's responses to all 60 scenarios in the session to train a logistic model, denoted by $H_{i,j} : \X\rightarrow \set{0,1}$, where $\X$ denotes the space of all pairwise comparisons.
Given any set of random comparisons $T$, let $p_{j_1, j_2}^{obs}(T)$ denote the fraction of $T$ where predictions from  $H_{i,j_1}$ and $H_{i,j_2}$ overlap, and let $p_{j_1, j_2}^{exp}(T)$ denote the fraction of overlaps \textit{expected by chance} if both models output independent Bernoulli trials with $p$ given by the empirical frequency of $1$ over $T$.
Then, \textit{model stability} between sessions $j_1$, $j_2$ is defined as\footnote{Note that the 
limit 
resolves the possibility of $\frac{0}{0}$ as $1$, which would otherwise make the expected value
undefined in most cases.}
\[
\text{MS}(j_1, j_2; i) := \E_{T}\left[ \lim_{\varepsilon \to 0^{+}} \frac{p_{j_1, j_2}^{obs}(T) - p_{j_1, j_2}^{exp}(T) + \varepsilon}{1-p_{j_1, j_2}^{exp}(T) + \varepsilon}\right].
\]

This measure, similar to Cohen's $\kappa$ (\citeyear{cohen1960coefficient}), is inspired by the works of \citet{geirhos2020beyond} and \citet{xu2025measuring} on behavioral alignment, 
and
quantifies disagreement between session-wise models,
reflecting changes to the underlying decision processes.
While the expectation removes dependence on any particular sample, the quantity still depends on the \emph{sample size}, but it rapidly converges. %
Henceforth, we \emph{estimate} $\text{MS}(j_1$, $j_2)$ by sampling a dataset $T$ containing 10k uniform random pairwise comparisons, and generating predictions from $H_{i,j_1}$, $H_{i,j_2}$ over $T$ to compute agreement.
We also quantify the \textit{cumulative model stability} of participant $i$ as $\textrm{cumulMS}(i) := \sum_{j=2}^5 \textrm{MS}(j-1, j; i).$

\paragraph{Model Shift.}
Along with stability measures, we quantify changes in properties of models learned using their responses.
We again use participants' responses to all 60 scenarios per session to train a {logistic model} for that session and learn Shapley values to obtain feature importances.
For participant any $i$ and session $j$, let $\smash{W^{\text{shap}}_{i, j}} := [w^{\textrm{(1)}}_{\smash{i,j}}, \dots, w^{\textrm{(d)}}_{\smash{i,j}}]$ denote the feature importance vector learned using this participant's responses for session $j$.
Then, \textit{model shift} for participant $i$ between sessions $1$ and $j$ is defined as
$$\textrm{model-shift}(j; i) := \sum_{j'=2}^j\norm{\vphantom{W}\smash{W^{\text{shap}}_{i, j'{-}1} - W^{\text{shap}}_{i, j'}}}_2^2,$$
quantifying the cumulative change in relative use of individual features across consecutive sessions between 1 and $j$.

\begin{table}
\centering
\begin{tabular}{@{\extracolsep{5pt}}lc}
\toprule
\textit{Dependent variable:} & Response stability $\textrm{RS}(\cdot, \cdot)$ \\
\midrule
 Intercept & $0.967^{***}$  (0.025) \\
 Scenario difficulty & $-0.043^{***}$ (0.002) \\
 Model-entropy & $-0.117^{***}$ (0.014) \\
 Mean reaction time & $-0.081^{**}$  (0.032) \\
 Scenario Variance & 0.039$^{}$ (0.027) \\
 User Variance & 0.131$^{***}$ (0.023) \\
 \hline \\[-1.8ex]
 Observations & 2414 \\
 Residual Std. Error & 0.121 (df=2420) \\
 \midrule
\textit{Note:} & \multicolumn{1}{r}{$^{*}$p$<$0.1; $^{**}$p$<$0.05; $^{***}$p$<$0.01} \\
\end{tabular}
\caption{Regression coefficients (std. dev. via Fisher information in brackets) for mixed-effects model of $\text{RS}(\cdot)$ vs.\ scenario difficulty, model-entropy, and reaction time for repeated scenarios. Significant associations here 
show that response instability is related to user's deliberation process.}
\label{tbl:mixed_effects_response_stability}
\end{table}

\paragraph{Model Complexity.} 
We also assess complexity associated with any participant's decision-making model.
For instance, a participant who uses just one patient feature to make the kidney allocation decision could be considered to have a model of lower complexity than a participant who uses several more features.
We measure the complexity of participant $i$'s process in session $j$ by computing the \textit{entropy} of the feature importance vector $\smash{W^{\text{shap}}_{i, j}} := [w^{\textrm{(1)}}_{i,j}, \dots, w^{\textrm{(d)}}_{i,j}]$.
Formally, 
$$\text{model-entropy}(i, j) := -\sum_{k=1}^d \frac{\abs{w^{\textrm{(k)}}_{i,j}}}{\norm{W^{\textrm{shap}}_{i, j}}_1} \ln \left( \frac{\abs{w^{\textrm{(k)}}_{i,j}}}{\norm{W^{\textrm{shap}}_{i, j}}_1}\right).$$
Lower values of $\text{entropy}(i, j)$ imply fewer features used for decision making by participant $i$ in session $j$.

\paragraph{Scenario Difficulty.}
Finally, we quantify properties relevant at the time of choice, such as extent to which a participant finds a scenario difficult.
Using the Bradley-Terry (BT) approach \cite{bradley1952rank}, for any pairwise comparison $A$ vs $B$, we can estimate the approximate priority scores implicitly assigned to each of $A$ and $B$.
For participant $i$, let $\beta^{(i)}_{\smash{F}}$ denote the weight on feature $F$, for any $F{\in}$ \{\textit{number of dependents}, \textit{life years gained}, $\ldots$\}.
Then, assuming priority scores for each patient can be modeled as a weighted linear combination of the patient's feature values, taking a BT approach, we get
$$\ln \frac{\mathbb{P}(\text{choose }A; i)}{\mathbb{P}(\text{choose }B; i)} 
    = \sum_{F} \beta^{(i)}_{\smash{F}} \cdot (A_F - B_F).$$
Now, suppose we learn weight vectors $\smash{\hat{\beta}^{(i)}}$ for each participant using logistic regression over participant responses to non-repeated scenarios. 
Then, using $\smash{\hat{\beta}^{(i)}}$s, we can calculate an implicit score that quantifies
the (subjective) \textit{difficulty} of participant $i$ for any repeated scenario.
To do this, we compute the relative score that the participant gives to 
either patient, i.e.,
$\sum_{F} \smash{\hat{\beta}_{\smash{F}}^{(i)}}{\cdot}A_F$ for $A$, likewise for $B$.
Then, the \textit{difficulty} of $A$ vs.\ $B$ can be captured as,  
$$\text{difficulty}(A,B; i) := -\left| \sum_{F} \smash{\hat{\beta}_{F}^{(i)}} \cdot (A_F - B_F) \right|.$$
This quantity measures the distance from a linear decision boundary for choosing either option,
where a difficulty of $0$ lies on the decision boundary (no evidence to prefer $A$ or $B$), and values become more negative to represent increasing certainty (
according to the logistic model).

\section{Findings: Preference Instability} \label{sec:observations_instability}

\subsection{Response Instability} \label{sec:observations_response_instability}
For participants who completed at least three sessions ($N{=}404$), Figure~\ref{fig:agreement_and_stability} presents response stability $\textrm{RS}(\cdot, \cdot)$ statistics.
We can see that response stability distribution trends toward higher values for scenarios with fewer patient feature tradeoffs ($U_1, U_2$), while it is significantly lower for scenarios with more tradeoffs ($V_1, V_2, W_1, W_2$). 
A Kruskal-Wallis test indicates that the differences in response stability levels across scenarios are significant $(H(5){=}222.23, p{<}0.001)$.

\begin{figure}
    \centering
    \includegraphics[width=0.7\linewidth]{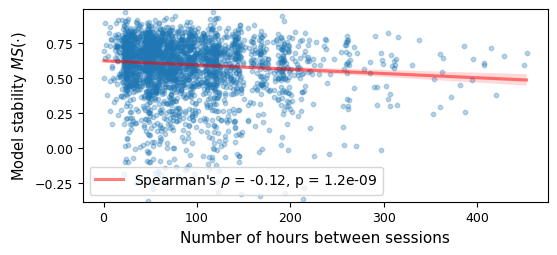}
    \caption{Model Stability between sessions $\textrm{MS}(\cdot, \cdot)$ vs.\ time difference between sessions. Significant negative correlation indicates decreasing model stability with time.}
    \label{fig:model_agreement_vs_time}
\end{figure}

We further assess possible reasons for participants' response instability.
Table~\ref{tbl:mixed_effects_response_stability} presents a mixed-effects model to predict response stability based on scenario difficulty, model entropy, and mean reaction time for the repeated scenarios. 
The results indicate a significant negative association between response stability and scenario difficulty, suggesting higher response instabilities for scenarios perceived as more difficult by the participants.
Similarly, there is a significant negative association between response stability and model-entropy, implying higher response instabilities for participants who use relatively more features in their decision process.
Finally, there is also a modest negative correlation between mean reaction time and stability, indicating possibly higher instability when participants deliberate on the scenario for a longer time.
We report detailed scenario-wise correlations between these measures in Appendix~\ref{sec:other_instability_results}.

All of these associations provide explanations for response instability; i.e. it is not just ``random noise'' or ``stochasticity,''  but rather associated with participant's mental processes, such as how difficult they found the scenario to be, how complex their decision-making process was, and how long they deliberated before making their decision.

\subsection{Model Instability}  \label{sec:observations_model_instability}
We also find evidence for shifts in decision-making models of participants over time.
Figure~\ref{fig:model_agreement_vs_time} provides evidence of this phenomenon.
We plot model stability between sessions $\text{MS}(j_1, j_2;i)$ of participant $i$ and sessions $j_1, j_2$ vs.\ the time difference between these sessions, for all $i, j_1, j_2$.
We observe a modest positive correlation between these quantities (Spearman's $\rho{=}-0.12, p{<}0.001$), suggesting an increase in disagreement between session-wise models over time.

In general, we hypothesize that different participants change their models in different ways.
While the above correlation may seem modest (though significant), further analysis indicates that some participants change their decision processes to a greater extent than others.
To discover mechanisms of preference change and their scale, we conduct a deep dive into different kinds of participants in our data.
\begin{figure}
    \centering
    \includegraphics[width=0.7\linewidth]{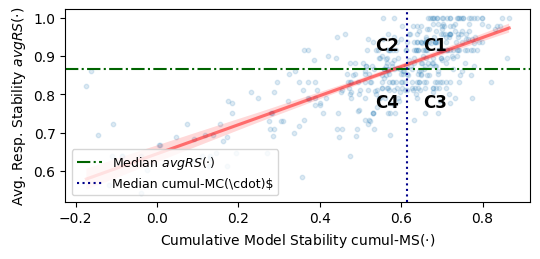}
    \caption{Average response stability vs.\ model stability, with participants categorized based on their plot location.}
    \label{fig:participant_categories}
\end{figure}

\subsection{Participant Categorization}  \label{sec:observations_categories}

We categorize participants by the cumulative \textit{response stability} and \textit{model stability} values, using groups based on these values to investigate mechanisms of preference change across the population.
For response stability, 
we divide the participant pool into \textit{response stable} vs.\ \textit{response unstable} groups along the median value of $\text{avgRS}(\cdot)$.
Similarly, for model stability, we
divide them into \textit{model stable} vs.\ \textit{model unstable} groups along the median value of $\text{cumulMS}(\cdot)$.
Based on these divisions, we get four participant categories: (C1) \textit{response stable, model stable}, (C2) \textit{response stable, model unstable}, (C3) \textit{response unstable, model stable}, and (C4) \textit{response unstable, model unstable}.
Figure~\ref{fig:participant_categories} visualizes these categories on the plot of response stability vs.\ model stability.
Note the significant positive correlation between response stability and model stability (Spearman's $\rho$=0.65, $p{<}.001$), leading to a larger number of participants in C1 (N=147) and C4 (N=153). In comparison, C2 (N=49) and C3 (N=55) have relatively fewer participants, but still sufficient to evaluate statistical differences across categories.

\begin{figure}
    \centering
    \includegraphics[width=0.7\linewidth]{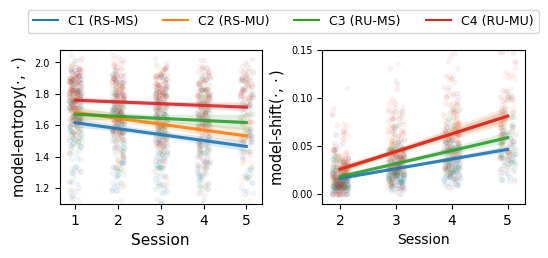}
    \subfloat[Session-wise entropy]{\hspace{.3\linewidth}}
    \subfloat[Model shift since session 1]{\hspace{.3\linewidth}}
    \caption{Session-wise model entropy and model shift for all categories. Participant categories differ in how their model properties change over time, revealing change mechanisms.}
    \label{fig:entropy_shift_by_session}
\end{figure}

With this categorization, we look at changes in model properties of participants over time to obtain deeper insight into their instabilities.
To that end, Figure~\ref{fig:entropy_shift_by_session} plots the change in model-entropy$(\cdot, \cdot)$ and
model-shift$(\cdot, \cdot)$ across sessions.
Based on these, we identify the following possible preference change mechanisms for different categories.

\textit{Mechanism 1: Increase the use of the ``most important'' feature over time.} 
For C1, Figure~\ref{fig:entropy_shift_by_session}a shows that these participants tend to start with a relatively low entropy model, i.e., their decision-making model is relatively simpler than others.
For C1 participants, model-entropy decreases across sessions ($r{=}-$0.23, $p{<}0.001$), suggesting that the preference change mechanism they tend to employ is to increase the use of features they consider most relevant.
Indeed, an analysis of the change in the Shapley value of the most important feature in Appendix~\ref{sec:other_instability_results} corroborates this finding.
Figure~\ref{fig:entropy_shift_by_session}a also shows that participants in C2 follow a similar mechanism for preference change ($r{=}{-}0.19, p=0.007$).
However, while the slopes are similar, 
C2 participants start with 
higher entropy than C1 on average.
Overall, participants in C1 and C2 tend to move toward lower entropy models over time.

\textit{Mechanism 2: Minimal preference change over time despite response instability.}
Participants in C3 are characterized by response instability and model stability.
Figure~\ref{fig:entropy_shift_by_session}a shows that model entropy for C3 is relatively high and stays high across sessions ($r{=}{-}0.08, p{=}0.2)$.
Taken together, this suggests that these participants tend to use a relatively larger set of features each session
and continue to use high entropy models (with minimal changes) across sessions.
Additionally, as negative correlation between model entropy and response instability in Table~\ref{tbl:mixed_effects_response_stability} shows, high entropy models possibly contribute to their response instability.

\textit{Mechanism 3: Significant updates to relative feature importances.}
Finally, participants in C4 are both response and model unstable.
Figure~\ref{fig:entropy_shift_by_session}a shows that these participants have the highest model entropy among everyone. Their model entropy remains high across sessions, suggesting they continue to use several features in the decision process.
Figure~\ref{fig:entropy_shift_by_session}b further shows that these participants exhibit high levels of model shift, significantly changing the importance they assign to different features over time ($r{=}0.74$, $p<0.001$).
This suggests that they may
not prioritize decision process consistency across scenarios (treating each scenario mostly independently) or are prone to relatively more \textit{stochastic} responses.
Participants in C4 also have shorter response times than C3 (Appendix~\ref{sec:other_instability_results}), suggesting shorter deliberations.

Overall, categorization by response and model instability sheds light on different potential sources of temporal instability in revealed preferences.
As we show next, this categorization has a strong bearing on AI alignment.

\begin{figure}
    \centering
    \includegraphics[width=0.7\linewidth]{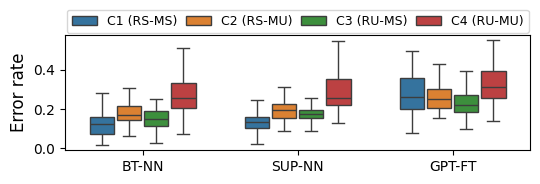}
    \caption{Error rate boxplot of all models, showing significant disparities in performance across participant categories.}
    \label{fig:aggregated_error_rate}
\end{figure}

\section{Findings: Impact on AI Alignment} \label{sec:observations_alignment}

With an understanding of how participants' preferences evolve over time, we next assess the impacts of preference evolution on
AI models learned from participant responses.
Here, we report the performance of three common preference modeling methods: (a) Bradley-Terry framework using neural scoring function (BT-NN), (b) Supervised neural networks (SUP-NN), (c) Fine-tuned GPT-2 model (GPT-FT).
BT-NN adapts the popular Direct Preference Optimization alignment framework \cite{rafailov2023direct} for this structured setting, while SUP-NN implements classic supervised preference learning.
We follow \citet{sun2024rethinking}'s methodology to train BT-NN and SUP-NN, and \citet{dickerson2025gets}'s process for the fine-tuned GPT.
Complete implementation details 
are presented in Appendix~\ref{sec:other_alignment_results}.

\paragraph{Aggregate error rate.}
We first assess aggregate model performance.
BT-NN and SUP-NN were participant-specific models, trained using each participant's data and evaluated over their held-out data 
(80-20 train-test split).
GPT-FT was fine-tuned by pooling together 50\% of all participant data and evaluated over held-out data from each participant.

Figure~\ref{fig:aggregated_error_rate} shows the performance of these models on participant-level data.
For all methods, we observe significantly higher error rates for response and/or model unstable participants (t-tests used to check significance).
For BT-NN, error rate is 0.16 higher on average 
for C4 participants compared to C1 participants $(t(290){=}14.5, p{<}0.001)$.
Similar trends hold for SUP-NN.
For GPT-FT, error rate is 
0.05
higher on average 
for C4 participants compared to C1 participants $(t(289){=}4.2, p{<}0.001)$.
However, GPT-FT has a higher error rate for all categories compared to BT-NN and SUP-NN, showing the limitations of population-level fine-tuning in aligning to individual preferences. 
Differences between mean error rates for C2 and C3 were not significant for BT-NN and SUP-NN, but error rates for C4 were 0.11
higher on average than that of C2/C3 for both BT-NN $(t(253){=}9.1, p{<}0.001)$ and SUP-NN $(t(253){=}11.0, p{<}0.001)$.
Error rates for C4 were also 0.08
higher on average than that for C2/C3 for both GPT-FT $(t(258){=}7.8, p{<}0.001)$.
Hence, performance of all methods is significantly worse for C4, and generally the error rate seems to increase with the level of instability.
Surprisingly, error rates for C1 were 0.03
higher on average than that for C2/C3 for GPT-FT $(t(253){=}2.7, p{=}0.008)$. 
This is likely because GPT-FT is fine-tuned over population data, leading it to perform generally worse in predicting individual responses.

\begin{figure}
    \centering
    \includegraphics[width=0.7\linewidth]{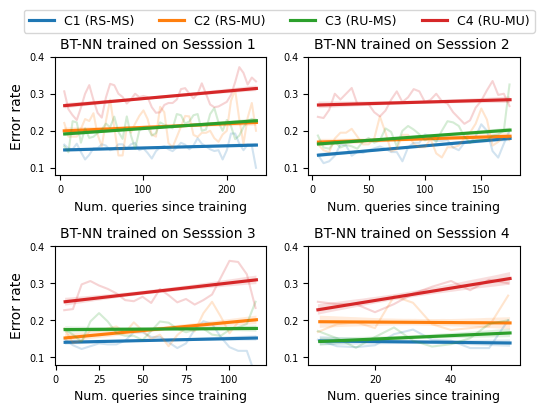}
    \caption{Error rate vs number of queries since training. The slope is positive and significant for C4 in 3 out of 4 settings.}
    \label{fig:error_rate_over_time}
\end{figure}

\paragraph{Error rate over time.}
Given temporal preference instability, we also investigate whether the AI models get worse over time.
For an AI model trained
on responses up to
time $t$, if the underlying preferences changed after $t$, 
then the model's error at time $t'{>}t$ 
increases with $t'$.
We first test this hypothesis using BT-NN model.
To test this hypothesis, we learn BT-NN model for each participant using their data from session $j$, and evaluate its error rate for sessions $j{+}1$ onward.
To study performance variation over time, we evaluate the error rate of the learned policy over batches of 10 consecutive queries, and assess trends in error rate as a function of the number of queries since training queries.

Figure~\ref{fig:error_rate_over_time} presents the results for this analysis.
The plots present the performance of participant-level models trained on each session.
For C4, the slope of error rate on number of queries since training is positive and significant when training is over session 1 (t=7.2, $p<0.001$), session 3 (t=6.2, $p<0.001$), and session 4 (t=5.9, $p<0.001$) responses.
It is, however, marginally non-significant when training over session 2 (t=1.8, p=0.06).
For these participants, AI performance gets worse over time in three out of four settings.
The magnitude of slope increase is relatively smaller for other participants, and reported in detail in Appendix~\ref{sec:other_alignment_results}.

\section{Discussion, Limitations, and Future Work} \label{sec:discussion}

Our findings highlight the challenges associated with the temporal instability of moral preferences.
While instability might appear to be
\textit{irrational behavior} on the surface, our observations suggest significant heterogeneity in preference change mechanisms, some of which may reflect \textit{authentic} updates to moral reasoning. 
This heterogeneity impacts AI alignment, as we see in the significant AI performance disparity across participant categories.
We next discuss the normative and technical implications of our findings.

\paragraph{What should we align with?}
This work raises normative questions at the core of AI alignment objectives. If moral preferences change significantly over time, \emph{to which preference should the AI be aligned?}
We could choose to align with moral preferences revealed during the latest time period, preferences revealed during an earlier time period, or 
some sophisticated combination thereof.
The choice between these candidates to serve as the objective of AI alignment is nontrivial, and depends on the reasons for observed differences between various candidate preferences. 

The implications of
this choice
are significant.
Aligning an AI to the ``ideal'' preference that a user thinks is normatively suitable would enhance trust placed in AI's decisions.
As \citet{jacovi2021formalizing} claim, ``{only when (1) the user successfully comprehends the true reasoning process of the model, and (2) the reasoning process of the model matches the user’s priors of agreeable reasoning, intrinsic trust is gained}''.
To this end, we argue that misalignment due to not capturing ``legitimate'' changes in moral preferences is bound to reduce the intrinsic trust the user places in AI.

\paragraph{Which preference changes are relevant to AI alignment?}
There are ``{legitimate}'' cases of preference change that should ideally be accounted for during AI alignment, while in 
``illegitimate'' cases, we might want to ignore them.
For instance, consider the gradual model shift exhibited by 
groups
C1 and C2, who reduced their model entropy over time and used fewer features in later sessions.
This shift could be a result of \textit{mental fatigue}, where the decision maker focuses only on the features they consider most important as a way to reduce cognitive load, time on task, or uncertainty \cite{jia2022effects}.
Depending on the context, one could argue that we should not account for this change during AI alignment, since we 
want to align an AI to the user's preferences for scenarios where they deliberated over all the presented information.
However, gradual model shift of C1 and C2 could also be due to legitimate reasons related to the nuances of \textit{preference construction}.
Preferences are known to be context-sensitive and largely shaped during decision-making \cite{warren2011values, slovic1995construction}.
People may start with complex decision models, but simplify them as they make more choices \cite{hoeffler1999constructing}.
If this is 
the mechanism driving the moral preference change for C1 and C2, then we should ideally
consider it
during AI alignment, as their later decisions reflect their well-formed preferences.

In contrast, 
groups C3 and C4 seem to use 
high-entropy models
across all sessions.
This could be because their preferences are not well-formed yet, in which case we need to continue presenting them with additional scenarios and align the AI to future well-formed preferences.
However, they might instead favor complex deliberations in each scenario (given the stakes of kidney allocation), in which case we could align the AI to their current preferences, while highlighting to them the inconsistencies in their responses to repeated scenarios. 
In 
either case, alignment requires understanding the mechanisms for preference change.
\ccolor{Yet, as we show, mechanisms are difficult to decipher at an individual level, even with longitudinal choice data, due to missing information on \textit{why} one changed their preferences. This appears to be a fundamental limitation of choice-based preference learning, 
necessitating richer reasoning-based data.
}

\paragraph{Technical solutions to address preference change.}
Depending on how one expects preferences to change, different solutions can be employed.
For participants who exhibited instability because they were learning their preferences during decision-making, \textit{assistive frameworks} model preference learning dynamics to ensure alignment with eventual well-formed preferences \cite{chan2019assistive, tian2023towards}.
Other works discount certain preferences. 
\citet{son2024right} assign higher weight to the latest choices during preference optimization.
\citet{zhuang2020consequences} argue for dynamic updates to alignment objectives when we have incomplete utility representations.
\citet{chowdhury2024provably} provide preference optimization methods that are robust to noisy human feedback.
However, the applicability of these works depends on 
the knowledge of preference change mechanisms.
Reweighting methods could be appropriate for legitimate preference changes,
while noise-robust optimization would be suitable for handling preference \textit{stochasticity}.
Yet, technical works on optimizing to dynamic preferences do not 
differentiate between various kinds of preference changes, a variety of which we observe to be potentially present in our single kidney allocation dataset.
As such, there is still a need for methodologies to understand how different people change their preferences and how to ensure alignment for all across time.

\paragraph{Opportunities beyond choice-based preference learning.}
While collecting user choices over an extended period allowed us to quantify temporal preference instabilities, we need richer data to decipher why participants changed their preferences in the manner we observed.
As such, future elicitation methods must go beyond simply learning from observed behavior.
Additional feedback from the users on alignment objectives could provide valuable signals to differentiate between competing objectives when faced with unstable preferences.
This feedback could be in the form of descriptions of ``evaluative concepts'' that serve as reasons behind our actions \cite{zhi-xuan_beyond_2024} or via ``{interactive alignment}'' frameworks that seek user input on desired AI goals, processes, and output assessments \cite{terry2023interactive}.

\paragraph{Limitations and Future Directions.} Our study was limited in certain ways that can be addressed in the future.
We assessed preference change over the span of days. However, one might expect larger preference changes over longer periods, which could be a fruitful target for future data collection efforts.
Our study also posed hypothetical kidney allocation decisions to laypeople, which may differ from decisions of medical professionals. 
Future work can also study other alignment objectives beyond prediction,  especially for language models \cite{carroll2024ai}.

\section*{Acknowledgments}
\ccolor{VK, CC, BKN, JSB, WSA, are grateful for the financial support from OpenAI and Duke University.

HH acknowledges support from the CMU-NIST Cooperative Research Center on AI Measurement Science \& Engineering (AIMSEC), and the AI Research Institutes Program funded by the National Science Foundation under AI Institute for Societal Decision Making (AI-SDM), Award No. 2229881. Any opinions, findings, conclusions, or recommendations expressed in this material are those of the authors and do not reflect the views of the funding agencies.

WSA discloses that he is owner and founder of Patient Preference Predictors, Inc., and a member of the Grow Therapy AI-Advisory Panel.}

\bibliography{references}

\counterwithin{figure}{section}
\counterwithin{table}{section}
\appendix

\clearpage

\section{Other Related Work} \label{sec:expanded_related_works}

\begin{figure}
\centering
    \fbox{\includegraphics[width=0.7\linewidth]{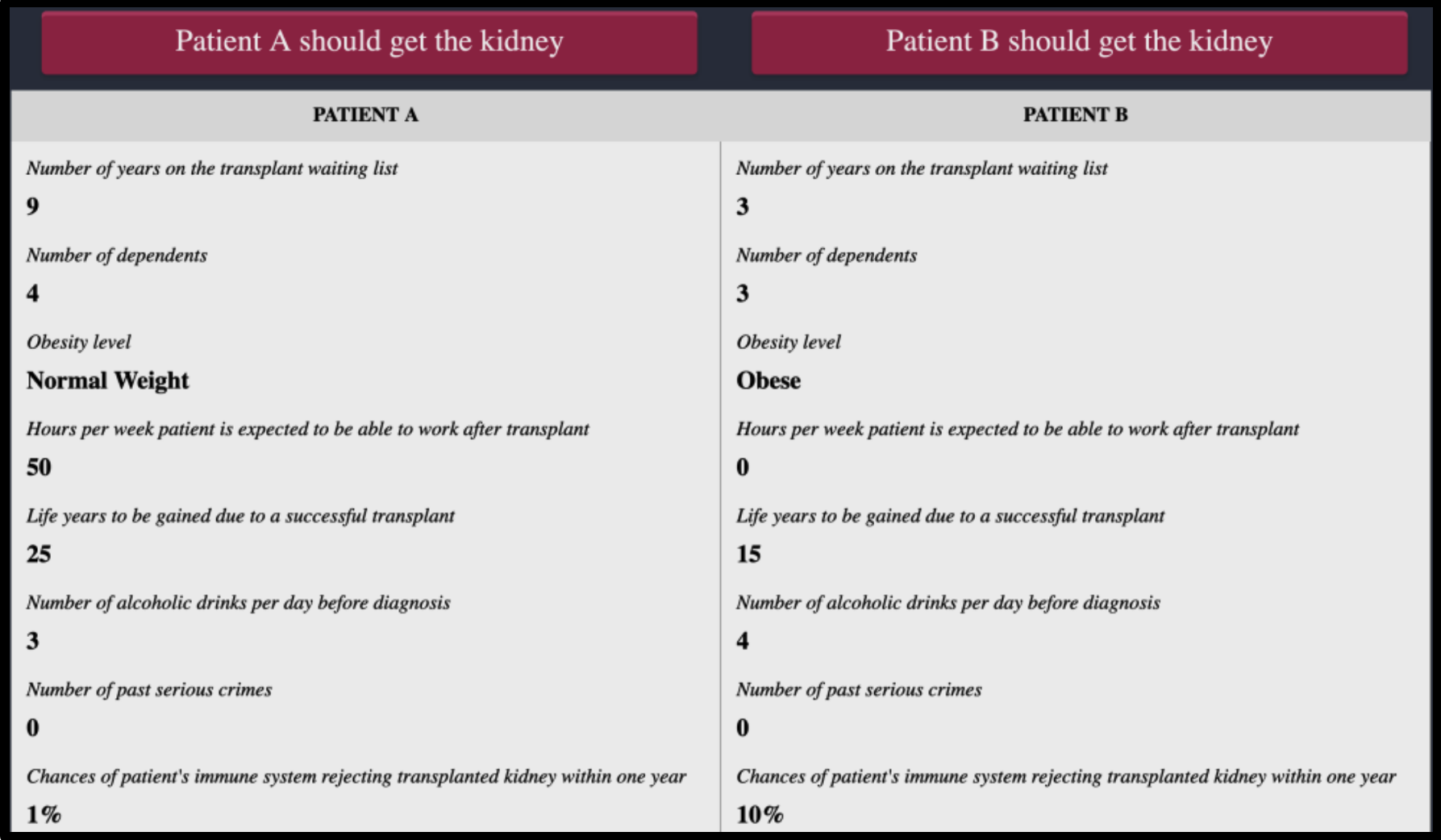}}
    \caption{Example of a pairwise comparison between hypothetical kidney patients.}
    \label{fig:example}
\end{figure}

\begin{table*}[]
    \centering
    \footnotesize
    \begin{tabular}{l | cc | cc | cc | cc | cc | cc | cc | cc}
    \toprule
        & \multicolumn{2}{c}{$U_1$} & \multicolumn{2}{c|}{$U_2$} & \multicolumn{2}{c}{$V_1$} & \multicolumn{2}{c|}{$V_2$} & \multicolumn{2}{c}{$W_1$} & \multicolumn{2}{c|}{$W_2$} & \multicolumn{2}{c}{$AT_1$} & \multicolumn{2}{c}{$AT_2$} \\
         Feature & A & B & A & B & A & B & A & B & A & B & A & B & A & B & A & B \\
    \midrule
         Number of dependents & 0 & 2 & 1 & 0 & 0 & 3 & 0 & 2 & 2 & 4 & 0 & 3 & 1 & 1 & 0 & 2  \\
         Life years gained & 15 & 10 & 20 & 10 & 10 & 10 & 10 & 10 & 10 & 15 & 10 & 5 & 15 & 0 & 0 & 20  \\
         Years on waiting list & 1 & 7 & 7 & 5 & 5 & 5 & 7 & 3 & 9 & 7 & 5 & 11 & 5 & 3 & 3 & 7  \\
         Weekly work hours post transplant & 20 & 30 & 30 & 40 & 20 & 20 & 20 & 20 & 40 & 30 & 30 & 25 & 40 & 0 & 0 & 30  \\
         Number of alcoholic drinks per day & 1 & 0 & 1 & 0 & 0 & 1 & 1 & 1 & 1 & 0 & 0 & 1 & 0 & 0 & 0 & 0  \\
         Number of past serious crimes & 2 & 0 & 0 & 2 & 0 & 2 & 0 & 0 & 0 & 2 & 2 & 1 & 0 & 1 & 0 & 0  \\
         Obesity level & 4 & 1 & 2 & 4 & 3 & 1 & 1 & 0 & 1 & 0 & 1 & 4 & 1 & 1 & 0 & 1  \\
         Chance 
            to reject transplant & 10 & 3 & 10 & 15 & 5 & 5 & 5 & 2 & 2 & 1 & 5 & 20 & 0 & 100 & 100 & 0  \\
    \bottomrule
    \end{tabular}
    \caption{All repeated scenarios, including the six scenarios $U_1, U_2, V_1, V_2, W_1, W_2$ repeated across all sessions and twice per session to measure response stability, and the two attention check scenarios $AT_1, AT_2$. For obesity, the following textual obesity values were provided -- 0: Underweight, 1: Normal weight, 2: Overweight, 3: Obese, 4: Morbidly obese.}
    \label{tab:repeated_scenarios}
\end{table*}

\begin{figure}
    \centering
    \includegraphics[width=0.7\linewidth]{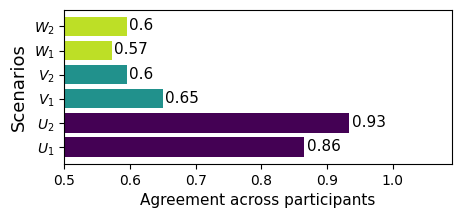}
    \caption{Agreement levels for all repeated scenarios across all participants.}
    \label{fig:agreement}
\end{figure}

AI alignment aims to ensure that AI systems behave in accordance with human values. A central challenge is that human preferences are not fixed; they may shift over time, introducing a distribution shift that undermines standard preference optimization approaches. 

One approach, Reinforcement Learning from Human Feedback (RLHF), aligns model outputs to human preferences by \emph{learning human preferences} from comparisons between alternative responses, then \emph{training AI systems} to optimize learned preferences \cite{christiano2017deep,ziegler2019fine,bai2022training}. This method creates an explicit reward model trained on a static dataset and then held constant during policy optimization. As \citet{carroll2024ai} show via simulations, when real-world preferences shift over time, a fixed reward model can cause misalignment phenomena such as reward hacking or goal drift.
In contrast, Direct Preference Optimization (DPO), removes the explicit reward model and optimizes the policy parameters directly from preference-labeled data, implicitly encoding preferences via supervised learning \citep{liu_survey_2025}. However, \citet{lin_limited_2024} find that DPO-trained models (and their implicit reward models) generalize poorly under distributional shifts, performing notably worse than explicitly-modeled reward systems when tested on out-of-distribution data. Thus, while RLHF and DPO differ in preference modeling, both share the same vulnerability: they assume preferences remain stable. Our work highlights the limitations of such assumptions as well.

Beyond this issue, alignment methods often aggregate heterogeneous human preferences into a single reward signal, assuming underlying homogeneity. In reality, values vary substantially between individuals and even \textit{within} individuals when considering temporal dynamics. Consolidating values across time and across populations into a
singular static objective
can yield inconsistent or unstable outcomes \citep{shirali_direct_2025}. Philosophical critiques further challenge the preferentist paradigm underlying both RLHF and DPO, emphasizing that values are dynamic, contextual, and not easily reducible to scalar quantities \citep{zhi-xuan_beyond_2024}. Empirical work supports this concern: even under stable conditions, individual judgments can fluctuate across time, raising doubts about the reliability of preference elicitation from single-session data \citep{boerstler2024stability}.

A growing body of research has considered using human moral preference data, often binary preference choices, to evaluate the alignment of AI models to human preferences \cite{sinnott2021ai,conitzer2024should}. For example, the widely cited Moral Machine dataset, which presents pairwise dilemmas regarding autonomous vehicle brake failure, has served as a benchmark for algorithmic voting systems \citep{noothigattu_voting-based_2018}, hierarchical Bayesian inference \citep{wiedeman_modeling_2020, kim2018computational}, and LLMs \citep{zaim_bin_ahmad_large-scale_2025}. Similarly, the ETHICS dataset, spanning five moral reasoning subtasks with two possible judgments, has become a baseline for evaluating moral reasoning in LLMs \citep{rodionov_evaluation_2023, hendrycks_aligning_2023}. This preference elicitation strategy spans even further into the domains of medical ethics \citep{dickerson2025gets, sinnott2021ai, freedman2020adapting}, resource triage \citep{mohsin_learning_2025}, algorithmic resource allocation \citep{johnston2023deploying,johnston2020preference,lee2019webuildai}, and multiwinner elections \cite{evequoz2022diverse}.  

The assumption that human moral preferences are a stable target also affects the many real-world applications of AI alignment. Much like theoretical approaches, applied alignment methods often treat human feedback as static, even when deployed in complex, real-world environments. In healthcare, preference data and clinician input have been used to inform ethically sensitive decisions such as treatment prioritization strategies \citep{van_leersum_human_2025}. Human input has also been used to train models to recognize when they should defer to human judgment, particularly in collaborative decision-making settings \citep{noti_learning_2023}. While these applications reflect a growing emphasis on human-centered alignment, they often rely on static snapshots of user preferences, overlooking how such preferences might shift over time.

While these applications and studies have proven valuable for investigations into alignment, they rely on moral judgments captured at one point in time and treating them as stable, generalizable targets. However, this assumption may be problematic if human moral preferences are themselves unstable over time and what is actually being captured is a single-session elicitation of preferences \cite{rehren2023stable,boerstler2024stability,warren2011values}. By contrast, longitudinal data can reveal if and how preferences change over time. Despite being framed as interactive or adaptive to human values, most current systems fail to incorporate such temporal tracking. If an individual's decision-making process can shift across time, models trained on static data risk learning preferences fail to generalize to future decisions.

\section{Additional Methodological Details} \label{sec:methodology_appendix}
In this section, we present additional details on study methodology that were excluded from the main body. Figure~\ref{fig:example} presents and example pairwise comparison.

\subsection{Feature Descriptions}
The following descriptions were presented to the participants about the patient features during onboarding.

a. Number of child dependents: The number of children who depend on the patient.
 
b. Number of alcoholic drinks per day before diagnosis: The patient’s drinking habits before being diagnosed with kidney failure.
 
c. Life years to be gained due to a successful kidney transplant: This is their gain in life expectancy if they receive the kidney transplant. For instance, suppose a particular patient would live 7 years without a kidney transplant or 15 years with a kidney transplant. The life years gained for this patient is then 8 years.
On average, patients with kidney disease are expected to live about 5 years without a kidney transplant.
 
d. Number of past violent crimes: The number of past serious violent crimes for which the patient was convicted, e.g., murder, aggravated assault, robbery, sexual assault, etc. For instance, a patient with a value of 2 for this feature could indicate that the patient has been convicted of committing two separate crimes of this nature in the past.
 
e. Obesity level: Whether the patient is underweight, normal weight, overweight, obese, or severely obese.
 
f. Hours per week patient is expected to be able to work after kidney transplant: Number of weekly work hours of the patient after a successful kidney transplant.

g. Years on the transplant waiting list: Number of years the patient has been on the waiting list for a kidney transplant.

h. 
Chance to reject transplant: Percent chance that the patient's immune system will reject the transplanted kidney in the near future.

\subsection{Descriptions of Repeated Scenarios}
The six pairwise comparisons that were repeated to test for response stability are presented in Table~\ref{tab:repeated_scenarios}.
The feature presentation order and the order of A/B was randomized every time the repeated scenario was presented.
The table also presents the attention check questions presented to the participants. Note the attention check questions were meant to be ``easy'' to answer (since one of A/B in these scenarios has 0 life years gained and 100\% chance of rejecting the transplant). Sessions where participants incorrectly answered any of the attention check questions were excluded from our analysis.

\subsection{End-of-survey Questions}
Participants who completed all five sessions were asked following additional questions at the end to obtain self-reported information on their decision process. 
\begin{itemize}
    \item ``What strategy or strategies did you use to decide which patient should be given the kidney in each scenario?''
    \item ``In scenarios where you found it difficult to decide, in particular, how did you ultimately determine who you thought should be given the kidney?''
    \item ``Please indicate to what extent did the relative increase in the value of each feature influence your decision to choose a patient for transplant'' (7-point scale presented for each feature with values ranging from ``strongly against'' to ``strongly in favor of'')
    \item ``Please rank the features based on how much influence they had on your decisions about who should get the kidney in each scenario, with rank 1 = the feature that had the most influence, and rank 8 = the feature that had the least influence'' (with ranks adjusted using a drag-and-drop tool)
\end{itemize}

Differences between self-reported and revealed preferences are discussed in Section~\ref{sec:other_instability_results}.

\subsection{Additional Dataset Details}

Demographic distribution of the overall participant pool along self-reported age, race, and gender was as follows:
\begin{itemize}
    \item Age: 31\% b/w 18--30, 51\% b/w 31--50, and 16\% 50+;
    \item Race: 74\% White, 13\% Black, 4\% Asian, and 9\% Other;
    \item Gender: 61\% Female, 38\% Male, 1\% Other.
\end{itemize}
Data on these three attributes were collected from all participants at the beginning of the first session to ensure the participant pool contained a sufficient number from each subgroup.
Beyond these attributes, we also asked participants to provide additional demographic information at the end of the fifth session.
Note that since not all participants completed all five days, this additional demographic data was available only for a subset of the overall participant pool.
For this subset, the distributions of values are noted below.
\begin{itemize}
    \item Education level: 22\% high-school graduate, 21\% some college, 20\% bachelor's degree, 14\% associate degree, 13\% master's degree, 4\% doctorate, 6\% other;
    \item Employment: 53\% employed full-time, 15\% homemaker, 10\% part-time, 5\% out of work, 4\% unable to work, 4\% student, 4\% self-employed, 2\% retired;
    \item Religion: 58\% Christianity, 2\% Buddhism, 2\% Islam, 2\% Judaism, 23\% Other; 13\% prefer not to answer;
    \item Social political orientation (0 to 8, ranging from ``extremely liberal'' to ``extremely conservative''): 4.4 $\pm$ 2.2;
    \item Economic and fiscal political orientation (0 to 8, ranging from ``extremely liberal'' to ``extremely conservative''): 4.5 $\pm$ 2.1.
\end{itemize}

The average time difference between consecutive sessions was around 45 hours. 
Among participants who completed all five sessions, the average overall time taken from the start of the first session to the end of the last session was around 201 hours.
Similarly, the average time between the start of the first session to the end of the fourth session was 140 hours, and the average time between the start of the first session to the end of the third session was 92 hours.
Mean time to complete each session was around 22 minutes.

\subsection{Preprocessing}
We remove responses where the participant's response time was more than two standard deviations away from the mean.
Additionally, since response time distributions can differ across participants, we standardize each participant's response time by subtracting their minimum response time and dividing by the difference between maximum and minimum response time, ensuring that the values lie between 0 and 1.

\begin{figure}
    \centering
    \includegraphics[width=\linewidth]{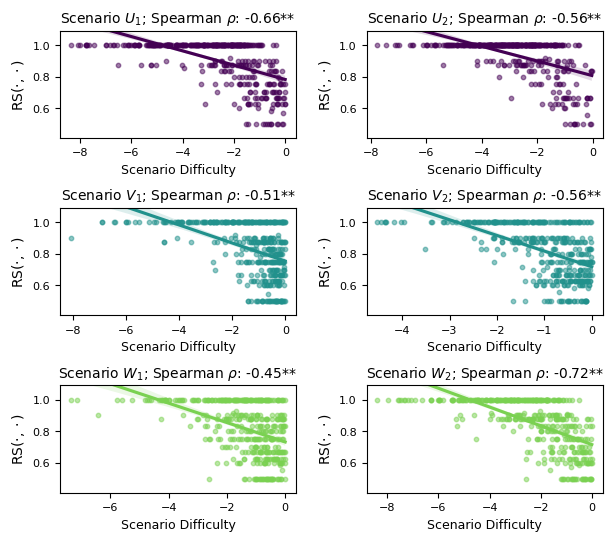}
    \caption{Response stability vs scenario difficulty for all repeated scenarios. Significant negative correlation ($**$ indicates $p{<}0.05$) for all scenarios shows that response stability is strongly associated with more difficult scenarios.}
    \label{fig:scenario_difficulty_stability}
\end{figure}

\begin{figure}
    \centering
    \includegraphics[width=\linewidth]{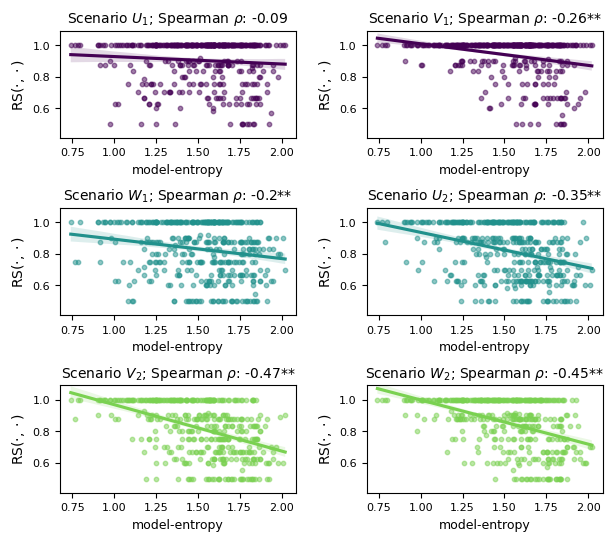}
    \caption{Response stability vs model entropy for all repeated scenarios. Significant negative correlation ($**$ indicates $p{<}0.05$) for most scenarios shows that response stability is associated with more complex decision models.}
    \label{fig:scenario_entropy}
\end{figure}

\subsection{Pilot Study}
We conducted a pilot to select appropriate scenarios to repeat in the main study.
The pilot consisted of a single session, and participants were recruited using Prolific.
67 participants took part in the pilot.

The pilot contained 17 candidate scenarios to repeat for the main study.
Of these, 5 had the fewest tradeoffs (with one feature favoring A/B and seven favoring the other), 6 had two tradeoffs (with two features favoring A, two favoring B, and the rest the same across patients), and the remaining 6 had four tradeoffs (with four features favoring A and four favoring B).
We checked how often participants agreed with each other in their response to these scenarios.
Based on agreement levels, we selected the scenarios that showed the widest range of agreement levels and that were qualitatively different from each other.

We use agreement to choose scenarios for response stability, since past work showed a significant association between across-participant agreement and within-participant stability \cite{boerstler2024stability}.
Indeed, Figure~\ref{fig:agreement} shows a similar association in our main study as well.

\section{Other Results on Preference Instability} \label{sec:other_instability_results}

In this section, we present additional results on preference instability.
Specifically, we expand on correlations between decision properties and response stability, as well as other differences between participant categories.

\subsection{Factors Associated With Response Instability}
Table~\ref{tbl:mixed_effects_response_stability} presents the regression analysis of response stability vs scenario difficulty and model entropy. 
We present additional plots for this analysis to show regressions between each repeated scenario and these factors.
Figure~\ref{fig:scenario_difficulty_stability} presents response stability vs scenario difficulty for all repeated scenarios and 
Figure~\ref{fig:scenario_entropy} presents response stability vs model-entropy for all repeated scenarios.

Additionally, we also find much lower across-participant agreement for scenarios where response stability levels were lower, as shown in Figure~\ref{fig:agreement}.

\subsection{Differences between Participant Categories}
We previously demonstrated differences in participant categories based on model entropy and model shift.
Here, we present additional differences related to the above to illustrate how participants in different categories differ from one another.

\begin{figure}
    \centering
    \includegraphics[width=0.6\linewidth]{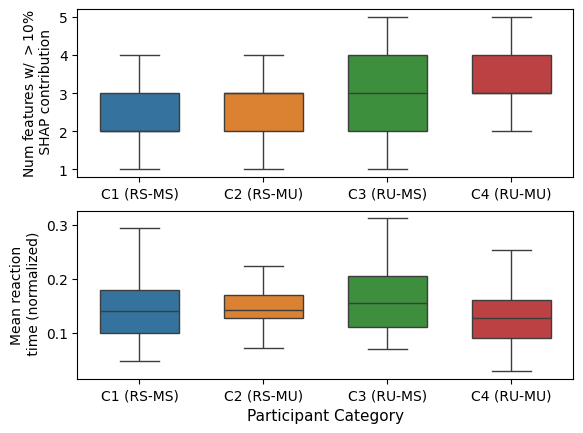}
    \caption{Number of features used and mean reaction time distributions for all participant categories}
    \label{fig:feature_usage_reaction_time}
\end{figure}

Figure~\ref{fig:feature_usage_reaction_time} shows distributions of the number of features used and the mean reaction time for all participant categories.
The number of features used approximates the size of the feature subset that was the most relevant to the participant's decision process.
To that end, we count the number of features for which the relative SHAP value was greater than $10\%$ in the learned logistic model. 

For both boxplots, we see significant differences across categories.
Participants in C4 use most features, followed by C3, and then C1/C2. This also aligns with the model-entropy levels for these categories, reported in Figure~\ref{fig:entropy_shift_by_session}.
Participants in C4 also have shorter reaction times than C3, suggesting that C4 participants take a shorter time to deliberate per scenario on average.

\begin{figure}
    \centering
    \includegraphics[width=0.6\linewidth]{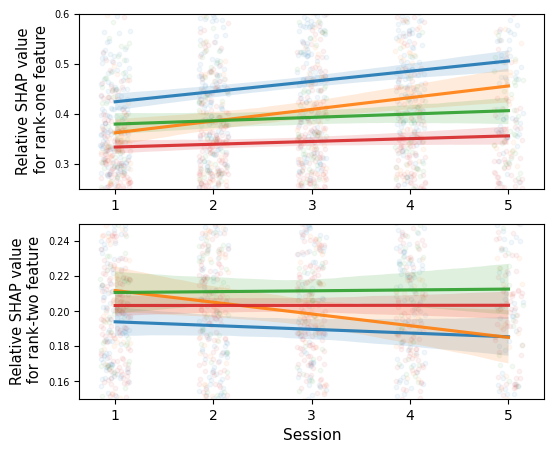}
    \caption{Relative importance of rank one and rank two feature per participant across sessions. Note that the rank one and rank two feature can differ across participants.}
    \label{fig:feature_usage_over_sessions}
\end{figure}

Figure~\ref{fig:feature_usage_over_sessions} plots the change in relative SHAP value for the participant's most important feature and second-most important feature across sessions.
We again see the mechanisms discussed in Section~\ref{sec:observations_categories} becoming clear here.
C1 and C2 significantly increase the importance of the most important feature. C2 also decreases the importance of the second-most important feature.
For C3 and C4, the weight assigned to the most important and second-most important feature remains almost the same across sessions.

\subsection{Differences between Self-reported and Revealed Preferences}
\begin{figure}
    \centering
    \includegraphics[width=0.7\linewidth]{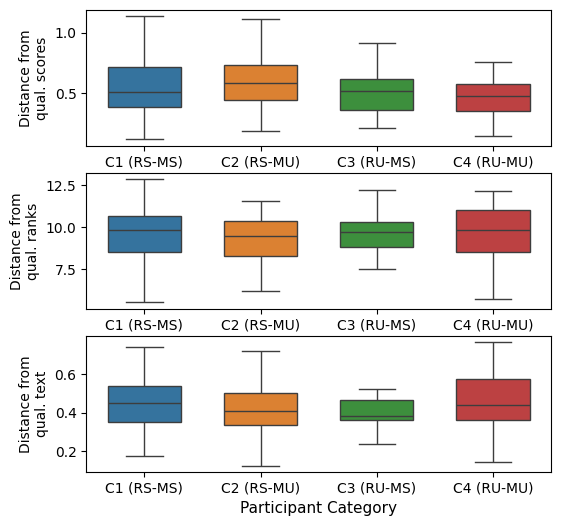}
    \caption{Distances between participants' self-reported preferences and revealed preferences (represented using feature SHAP distribution).}
    \label{fig:qualitative_distances}
\end{figure}

Figure~\ref{fig:qualitative_distances} presents various measures of distance between participants' self-reported preferences and revealed preferences.
The self-reported preferences included their self-reported importance scores for each feature, the rank of each feature, and indicator values for whether a feature was mentioned in their self-reported decision strategy.
The revealed preferences included their SHAP feature importance distribution.

\paragraph{Feature scores.} Participants provided scores on how much each feature influenced their decision process (from -3 to 3).
We normalize the provided absolute values of self-reported scores and compute the distance between these scores and the normalized SHAP feature values.
The top plot in Figure~\ref{fig:qualitative_distances} presents this measure for all participant categories.
Surprisingly, the distance between self-reported and SHAP feature importances is for the most response stable groups, C1 and C2. This is likely because, even though these participants use very few features in their decision process, they still self-report relatively more features as being relevant, leading to the observed discrepancy.
In comparison, C3 and C4 use many features in the decision process and also report several of those features to be important, leading to smaller differences between self-reported and SHAP feature importances.
A Kruskal-Wallis test indicates that the differences in distance between self-reported and SHAP feature importances across categories are significant ($H(3) = 7.69, p=0.05$).

\paragraph{Feature ranks.} Participants also provided feature ranks.
We compute revealed ranks from the SHAP feature distribution for each participant and report the difference between revealed and self-reported feature ranks in the middle plot of Figure~\ref{fig:qualitative_distances}.
Here, the differences between participant categories are relatively smaller.
A Kruskal-Wallis test indicates that the differences in distance between self-reported and SHAP feature ranks across categories are not significant ($H(3) = 1.63, p=0.65$).
Hence, participants' self-reported ranks were more consistent across categories with the feature ranks learned from their responses.

\paragraph{Text-based strategy response.} Participants were also optionally asked to describe their decision process in words.
We extract the features they mention in their responses and use those to create an indicator vector over the feature set, with 1 for features mentioned in their response and 0 for ones not mentioned.
Then, we compute the difference between the normalized indicator vector for the participant vs the normalized SHAP feature importance vector.
The distribution of this measure is reported in the bottom plot of Figure~\ref{fig:qualitative_distances}.
Once again, we find relatively smaller differences across groups for this measure.
A Kruskal-Wallis test indicates that the differences in this measure across categories are not significant ($H(3) = 4.11, p=0.25$).

\section{Other Details and Results on AI Alignment} \label{sec:other_alignment_results}

\subsection{Implementation Details}

\paragraph{BT-NN Implementation Details}
We implemented BT-NN using the Bradley-Terry framework and with a neural network backbone \cite{bradley1952rank}.
In particular, this model assigns \textit{priority scores} to each option in a pairwise comparison and selects the option with a higher score.
The scores are generated using a neural network with two hidden layers and ReLU activation. 
The model parameters are learned by minimizing the cross-entropy loss of the difference between scores for options A and B in any given pairwise comparison.
We use Adam for stochastic batch optimization, with a learning rate of 1e-3 and batch sizes of 16.
The implementation used the Python Keras library.

\paragraph{SUP-NN Implementation Details}

SUP-NN is implemented using the MLPClassifier function in the Python scikit-learn library.
We again use ReLU activation functions and Adam for stochastic batch optimization with a learning rate of 1e-3. 

\paragraph{GPT FT Implementation Details}
We fine-tuned a GPT-2 model for binary classification using the Hugging Face Transformers library. For each participant, 50\% of their responses were randomly selected and pooled into a global training dataset, while the remaining 50\% were used for testing. The final dataset was split into training (90\%) and validation (10\%) sets using a fixed random seed for reproducibility.
\begin{itemize}
    \item \textbf{Model and Tokenization}
    \begin{itemize}
        \item Base model: GPT-2 from Hugging Face Transformers.
        \item Tokenizer: AutoTokenizer from the GPT-2 checkpoint.
        \item Padding token was set to the end-of-sequence token (EOS).
        \item The model’s embedding layer was resized to match the tokenizer vocabulary.
        \item A classification head was added with two output labels (num\_labels = 2).
    \end{itemize}
    \item \textbf{Preprocessing}
    \begin{itemize}
        \item Difference between patient values were converted to text format.
        \item Text inputs were tokenized with a maximum sequence length of 128, using truncation and padding to fixed length.
        \item Labels were assigned during tokenization and stored in the “labels” field.
        \item Metadata columns (text, tid, session\_number, created\_at) were removed.
        \item All datasets were converted to PyTorch tensors using the fields: input\_ids, attention\_mask, and labels.
    \end{itemize}
    \item \textbf{Training Configuration}
    \begin{itemize}
        \item Training was conducted using the Hugging Face Trainer API with a PyTorch backend.
        \item Batch size: 8 per device
        \item Gradient accumulation steps: 2
        \item Number of training epochs: 3
        \item Weight decay: 0.01
        \item Random seed was set to 42 for reproducibility
    \end{itemize}
    \textbf{Hardware}
    \begin{itemize}
    \item Training and evaluation were performed on a Mac mini (M4) with:
    \begin{itemize}
        \item 12-core CPU
        \item 16-core GPU
        \item 16-core Neural Engine
        \item 64 GB unified memory
    \end{itemize}
    \end{itemize}
\end{itemize}

\subsection{Regression Results for Error Rate}
Figure~\ref{fig:error_rate_over_time} shows the error rate change over time for different participant categories.
The chosen method for participant categorization provides a discrete, interpretable way to assess the impact of response and model instabilities on temporal misalignment.
Here, we also provide a continuous assessment for this result by performing an OLS regression of error rate vs query number, average response stability avgRS$(\cdot)$, and cumulative model instability cumulMS$(\cdot)$.

Table~\ref{tbl:regression_bt_nn} presents the results of this regression for BT-NN models. The results expand on those presented in Figures~\ref{fig:aggregated_error_rate} and ~\ref{fig:error_rate_over_time}.
Each column presents the results for the model trained on a specific session between 1 and 4.
For all training sessions, we see that the coefficient of C4 is larger than other categories, confirming the finding of Figure~\ref{fig:aggregated_error_rate} that the error rate is much larger for C4 than for other categories in most cases, even when training on individual session data.
Additionally, we also see a statistically-significant positive coefficient associated with the log of query number for training sessions 1, 2, and 3, indicating that for the default category (i.e., C1), error rate increases with query number for these sessions.
To discover the coefficient for any other category $C$, we need to take the sum of the coefficient of log(query number) and that of the interaction term log(query number)$^\star\mathbf{1}[C]$.
With this computation, for C2, models trained on sessions 1, 2, or 3 also show an increase in error rate with log(query number).
For C4, models trained on all four sessions show an increase in error rate with log(query number).
Finally, for C4, models trained on all sessions 1, 2, and 4 show an increase in error rate with log(query number), as also observed in Figure~\ref{fig:error_rate_over_time},

Tables~\ref{tbl:regression_sup_nn} present the same regression analysis for error rates of SUP-NN models.
Overall, the trends of relative dependence of error rate on the participant categories remain similar here.
Although the coefficients for log(query number) and interaction terms between log(query number) and participant categories are slightly different for these models compared to BT-NN.
We do not run this analysis for GPT-FT since session-specific participant-level datasets are not large enough to obtain non-trivial accuracy for fine-tuned LLMs.

\begin{table*}[!htbp] \centering
\footnotesize
\begin{tabular}{@{\extracolsep{5pt}}lcccc}
\toprule
& \multicolumn{4}{c}{\textit{Dependent variable: BT-NN error rate}} \
\cr \cline{2-5}
\\[-1.8ex] & Session 1 model & Session 2 model & Session 3 model & Session 4 model \\
\midrule
 $\mathbf{1}$[C1 (RS-MS)] & 0.139$^{***}$ & 0.106$^{***}$ & 0.125$^{***}$ & 0.136$^{***}$ \\
& (0.008) & (0.009) & (0.011) & (0.016) \\
 $\mathbf{1}$[C2 (RS-MU)] & 0.232$^{***}$ & 0.164$^{***}$ & 0.092$^{***}$ & 0.216$^{***}$ \\
& (0.014) & (0.016) & (0.019) & (0.028) \\
 $\mathbf{1}$[C3 (RU-MS)] & 0.140$^{***}$ & 0.126$^{***}$ & 0.177$^{***}$ & 0.093$^{***}$ \\
& (0.009) & (0.010) & (0.012) & (0.018) \\
 $\mathbf{1}$[C4 (RU-MU)] & 0.243$^{***}$ & 0.265$^{***}$ & 0.235$^{***}$ & 0.195$^{***}$ \\
& (0.009) & (0.011) & (0.012) & (0.019) \\
 log(query number) & 0.006$^{***}$ & 0.011$^{***}$ & 0.005$^{**}$ & 0.002$^{}$ \\
& (0.002) & (0.002) & (0.003) & (0.005) \\
 log(query number)$^\star\mathbf{1}$[C2(RS-MU)] & -0.007$^{*}$ & -0.005$^{}$ & 0.020$^{***}$ & -0.007$^{}$ \\
& (0.003) & (0.004) & (0.006) & (0.010) \\
 log(query number)$^\star\mathbf{1}$[C3(RU-MS)] & 0.008$^{***}$ & 0.003$^{}$ & -0.008$^{*}$ & 0.020$^{***}$ \\
& (0.003) & (0.003) & (0.004) & (0.007) \\
 log(query number)$^\star\mathbf{1}$[C4(RU-MU)] & 0.006$^{**}$ & -0.008$^{**}$ & 0.006$^{}$ & 0.021$^{***}$ \\
& (0.003) & (0.003) & (0.004) & (0.007) \\
\hline \\[-1.8ex]
 Observations & 28219 & 20702 & 13530 & 6117 \\
 $R^2$ & 0.122 & 0.114 & 0.140 & 0.126 \\
 Adjusted $R^2$ & 0.121 & 0.114 & 0.140 & 0.125 \\
 Residual Std. Error & 0.140 & 0.139& 0.137 & 0.136 \\
 F Statistic & 558.301$^{***}$ & 380.065$^{***}$ & 314.415$^{***}$  & 126.083$^{***}$  \\
\hline
\hline \\[-1.8ex]
\textit{Note:} & \multicolumn{4}{r}{$^{*}$p$<$0.1; $^{**}$p$<$0.05; $^{***}$p$<$0.01} \\
\end{tabular}
\caption{OLS regression of error rate of BT-NN model vs cumulative response and model instabilities.}
\label{tbl:regression_bt_nn}
\end{table*}

\begin{table*}[!htbp] \centering
\footnotesize
\begin{tabular}{@{\extracolsep{5pt}}lcccc}
\toprule
& \multicolumn{4}{c}{\textit{Dependent variable: SUP-NN error rate}} \
\cr \cline{2-5}
\\[-1.8ex] & Session 1 model & Session 2 model & Session 3 model & Session 4 model \\
\midrule
 $\mathbf{1}$[C1 (RS-MS)]  & 0.213$^{***}$ & 0.114$^{***}$ & 0.132$^{***}$ & 0.144$^{***}$ \\
& (0.008) & (0.010) & (0.011) & (0.017) \\
 $\mathbf{1}$[C2 (RS-MU)]  & 0.212$^{***}$ & 0.227$^{***}$ & 0.164$^{***}$ & 0.187$^{***}$ \\
& (0.015) & (0.017) & (0.020) & (0.031) \\
 $\mathbf{1}$[C3 (RU-MS)]  & 0.219$^{***}$ & 0.157$^{***}$ & 0.178$^{***}$ & 0.153$^{***}$ \\
& (0.010) & (0.011) & (0.013) & (0.020) \\
 $\mathbf{1}$[C4 (RU-MU)]  & 0.278$^{***}$ & 0.268$^{***}$ & 0.294$^{***}$ & 0.178$^{***}$ \\
& (0.010) & (0.012) & (0.013) & (0.021) \\
 log(query number) & -0.002$^{}$ & 0.019$^{***}$ & 0.012$^{***}$ & 0.010$^{**}$ \\
& (0.002) & (0.002) & (0.003) & (0.005) \\
 log(query number)$^\star\mathbf{1}$[C2(RS-MU)]  & 0.011$^{***}$ & -0.017$^{***}$ & 0.000$^{}$ & -0.009$^{}$ \\
& (0.004) & (0.004) & (0.006) & (0.011) \\
 log(query number)$^\star\mathbf{1}$[C3(RU-MS)]  & 0.008$^{***}$ & -0.003$^{}$ & -0.004$^{}$ & 0.008$^{}$ \\
& (0.003) & (0.003) & (0.004) & (0.008) \\
 log(query number)$^\star\mathbf{1}$[C4(RU-MU)] & 0.013$^{***}$ & -0.007$^{**}$ & -0.005$^{}$ & 0.029$^{***}$ \\
& (0.003) & (0.003) & (0.004) & (0.008) \\
\hline \\[-1.8ex]
 Observations & 28219 & 20702 & 13530 & 6117 \\
 $R^2$ & 0.100 & 0.099 & 0.143 & 0.114 \\
 Adjusted $R^2$ & 0.100 & 0.099 & 0.143 & 0.113 \\
 Residual Std. Error & 0.147 & 0.152  & 0.141 & 0.150 \\
 F Statistic & 447.398$^{***}$ & 325.924$^{***}$ & 322.944$^{***}$ & 112.376$^{***}$  \\
\hline
\hline \\[-1.8ex]
\textit{Note:} & \multicolumn{4}{r}{$^{*}$p$<$0.1; $^{**}$p$<$0.05; $^{***}$p$<$0.01} \\
\end{tabular}
\caption{OLS regression of error rate of SUP-NN model vs cumulative response and model instabilities.}
\label{tbl:regression_sup_nn}
\end{table*}

\end{document}